# Control of 1,3-Cyclohexadiene Photoisomerization Using Light-Induced Conical Intersections

*Jaehee Kim\*, Hongli Tao, James L. White, Vladimir S. Petrović, Todd J. Martinez, and Philip H. Bucksbaum*

PULSE Institute, Stanford University, Stanford, CA 94304

\*Corresponding author: jk2236@stanford.edu

ABSTRACT.  We have studied the photo-induced isomerization from 1,3-cyclohexadiene to 1,3,5-hexatriene in the presence of an intense ultrafast laser pulse.  We find that the laser field maximally suppresses isomerization if it is both polarized parallel to the excitation dipole and present 50 fs after the initial photoabsorption, at the time when the system is expected to be in the vicinity of a conical intersection that mediates this structural transition.  A modified *ab initio* multiple spawning (AIMS) method shows that the laser induces a resonant coupling between the excited state and the ground state, i.e., a light-induced conical intersection.  The theory accounts for the timing and direction of the effect.

KEYWORDS  Quantum control, conical intersections, isomerization, cyclohexadiene, ab initio multiple spawning

## 1. Introduction

The Born-Oppenheimer (BO) approximation, which assumes separation of electronic and nuclear motion, underpins our current understanding of molecular structure.[1,2] The approximation accurately describes a molecular system near the equilibrium geometry. Yet, chemical transformations inherently require the molecule to leave this geometry, and the regions of potential energy surface (PES) where the BO approximation breaks down can be critical from the viewpoint of chemical reaction dynamics.[3-6] Conical intersections (CIs), locations where two potential energy surfaces in a polyatomic molecule become degenerate, are found to mediate most photochemical processes.[7-11] At conical intersections, the BO approximation breaks down and wavepacket motion on more than one potential energy surface needs to be considered.[12-17]

Intense external electromagnetic fields have the ability to couple different electronic states, thus modifying the topology of potential energy surfaces.[18,19] A wavepacket propagating on a field-dressed PES will experience different gradients, compared to the unmodified PES, potentially leading to a different mechanism and/or product. The coupling induced by the external field can be either resonant or nonresonant, and the relative contributions of the two depend on the field amplitude, frequency, polarization, and energy separation between the two PESs.[20-25]

The interaction of a coherent laser field with a molecule can be considered in a dressed state Born-Oppenheimer basis where the resonance is a point of crossing between two BO potential energy surfaces whose bare energies differ by one photon at that point. The dipole coupling then takes the form of off-diagonal matrix elements in the dressed BO Hamiltonian. This is often referred to as a Floquet picture,[26-30] and it permits us to consider the coupled electron-atom dynamics using the same formalism as BO dynamics. The laser polarization acts as an additional degree of freedom in the problem. In general, the off-diagonal field couplings cannot lift the degeneracy in all of the degrees of freedom, and therefore laser-induced conical intersections (LICIs) dominate the dynamics near electronic resonances.[31-35] In the usual (not light-induced) molecular CIs, the degeneracy is lifted (in first order) by two collective molecular coordinates.[36] In contrast, the laser polarization direction plays the role of one of these degeneracy-lifting coordinates in the context of LICIs.[31-35,37,38] Within a dressed state picture, the dynamics at LICIs can be modeled using similar methods as those used to describe nonadiabatic dynamics around avoided crossings or CIs.[21,39-43]

Ring-opening of 1,3-cyclohexadiene (CHD) is among the simplest photoinduced conrotatory electrocyclic reactions and it is a prototype for many important biochemical pathways, such as vitamin D production.[44,45] The photoinduced ring-opening of CHD is ultrafast (with ground state recovery within 200fs of photoexcitation) and thus it has been studied extensively with ultrafast spectroscopic techniques (see [46-50] and references therein). It has also been studied theoretically using both static electronic structure methods to characterize conical intersections[51-53] and quantum dynamics with reduced dimensionality models.[54-57] The isomerization process is initiated by absorption of a single UV photon, which excites a nuclear wave packet in the Franck-Condon region to a spectroscopically bright 1B state. The excited wavepacket rapidly crosses onto the 2A state, whence it continues to evolve towards a conical intersection with the 1A ground state. In this description the symmetry labels are only approximate, since while the isomerization process is underway the molecule no longer retains its initial $C_{2V}$ symmetry. The 2A/1A conical intersection geometry is very close to the transition state geometry for isomerization from CHD to 1,3,5-hexatriene (HT) on the ground electronic state. Thus, the wavepacket arrives to the ground state in a region where it is expected to branch nearly equally between the two possible outcomes (regeneration of the CHD reactant or formation of the HT photoproduct). This implies that rather small perturbations to the dynamics around this 2A/1A intersection could have a large impact on the branching ratio. We demonstrate here that this is indeed the case, using a timed intense-IR pulse to influence the ring-opening along the path to the 2A/1A CI and thereby suppressing the formation of the HT isomerization product.

These experiments require that we monitor the HT production as the control parameters are varied. We do this by means of laser-induced ionization and fragmentation of the molecule after it returns to the ground state. Using Principal Component Analysis (PCA), we are able to identify the different isomerization channels in the CHD to HT reaction through their unique fragmentation signatures.[49] This method has been used previously to show that that changes in $H^+$ yield are primarily due to changes in the number of HT molecules detected. In particular, by monitoring the $H^+$ count at a fixed delay one can monitor the extent of HT production.[47,58-60] Here we have used this method to investigate the effect of a 'control' pulse on the HT production.

We have also developed an explicit treatment of light-matter interactions within the *ab initio* multiple-spawning (AIMS) framework,[61,62] which propagates wavepacket dynamics in full-dimensionality simultaneously with solution of the electronic Schrödinger equation in order to obtain the potential energy surfaces, gradients, and nonadiabatic couplings "on the fly." We use this modified AIMS method to simulate the control experiment. The light-matter interaction is described in the dipole approximation for these calculations, where the required dipole matrix elements are also calculated "on the fly." We have also modified the original spawning algorithm that is used to adaptively increase the size of the nuclear basis set. While the usual AIMS method increases the size of the basis set in response to large molecular nonadiabatic coupling matrix elements, we now also "spawn" when the off-diagonal elements of the Hamiltonian that describe the light matter interaction are large. In this way, we can describe nonadiabatic dynamics around both normal and light-induced avoided crossings and CIs within the same theoretical and computational framework.

## 2. Experiment

A Ti:sapphire laser with duration of ~70 fs and center wavelength at 800 nm is divided into three pulses: the first for up-conversion to 266 nm to be used to initiate the ring opening, the second that is used as a control field to produce the light-induced conical intersection, and the third that is used to photofragment the molecule via Coulomb explosion to learn whether it has isomerized. The UV radiation was generated in two stages: frequency doubling in Type I BBO, followed by sum frequency generation in Type I BBO. The resulting UV pulse had 13 µJ with 120 fs FWHM centered at 266 nm. The intensity of the UV pulse was strong enough to produce a high fraction of molecules in the excited state. Its strength was adjusted so that approximately half of the CHD ions detected in the experiment were created by multiphoton excitation by the UV pulse alone, and half were produced in the subsequent Coulomb explosion. The two 800 nm pulses, one for control and one for fragmentation analysis, were combined at a beam splitter, and they were both combined with the UV pulse in a dichroic mirror to achieve a collinear propagation of the three pulses. In order to ensure that the fragmentation pulse probed molecules excited by the pump and influenced by the control pulse, the relative spot sizes at the focus were arranged in the following order: probe spot size < control pulse spot size < UV pump spot size. The spot sizes for the three beams in the interaction region were 100µm, 50µm, and 30µm for the UV, control, and probe pulses, respectively. The intensity of the control pulse (~23 µJ) was adjusted such that no appreciable fragmentation of the UV-excited sample was observed without the probe pulse. The fragmentation pulse energy (100 µJ) was selected for clear distinction between the CHD and HT fragmentation patterns, as described in our previous work.[49] We estimated the intensity of each pulse at the interaction region is: UV pump $\sim 10^{11}$ W/cm$^2$, IR control pulse $\sim 10^{12}$ W/cm$^2$, and IR probe pulse $\sim 10^{13}$ W/cm$^2$. The polarizations of the UV pump and the IR probe were parallel and the polarization of the IR control pulse was varied. The time delay of the two IR laser pulses with respect to the UV laser pulse could be controlled independently. The delay of the probe pulse with respect to the UV pulse was fixed at 50 ps. The delay of the control pulse with respect to the UV pulse was scanned in the range between -500 fs and +900 fs in 10 fs steps. The excitation scheme is shown in Fig. 1(a).

The 1,3-Cyclohexadiene source (Aldrich, 97%) was used without any purification. The source was an effusive room temperature beam (vapor pressure ~10 mbar) directed through a skimmer (2mm diameter) in a direction perpendicular to the propagation of the laser beams. We extracted positive ions from photofragmentation using a 750 V/cm field in a time-of-flight (TOF) mass spectrometer equipped with a 40 mm dual MCP (Jordan). On each laser shot the ions were detected with sufficient resolution to separate different mass peaks, almost all of which consisted of singly ionized $C_{X\leq 6}H_{Y\leq 8}$ fragments (1 to 80 amu). To reduce shot-to-shot noise in our data, traces from 500 shots were averaged at each delay.

### 3. Experimental Results

The main method used to analyze these mass spectra is principal component analysis (PCA). This is a covariance technique in which the characteristic fragmentation spectra for different product isomers are found by analyzing the common variance patterns in the data.[63] Disappearance of the parent molecule CHD and the appearance of the daughter molecule HT could be followed as a function of the relative delays and relative polarizations of the UV excitation pulse and the IR control pulse.

The main result of this work is that we observe a strong influence of the control pulse on the photofragmentation ion-TOF spectra. By adjusting the timing and polarization of the control pulse, we are able to suppress HT production. The maximum suppression of HT production is observed when the control pulse arrives approximately 50 fs after the photoexcitation. The magnitude of the suppression depends on the relative polarization between the photoexcitation pulse and the control pulse and is greatest when the two pulses are polarized parallel to each other. In the rest of this section we will describe the features of the data that led to this conclusion.

A multiphoton absorption-induced cross-correlation (temporal overlap) between the pump and the control pulses results in an increase of ion count when the two pulses arrive simultaneously, a condition that we call "time zero." All peaks in the ion-TOF mass spectrum display this increase. In addition to the cross correlation signal, selected lighter fragments showed a visible dip in their ion count at ~50 fs.

To extract the effect of the control field in presence of cross-correlation signal, we employed principal component analysis (PCA) as suggested by our previous work.[49] We find that more than 97% of the variance in the TOF spectra in the current experiment can be explained by two characteristic spectra corresponding to CHD and HT.[49] When we reduce the data set to two dimensions that adequately describe cross-correlation and HT production, we filter out noise and effects that have a minor contribution to the evolution of the fragmentation spectra in our experiment. We identify a basis to represent the two dimensional data, so that the in the new basis the coefficients for the two processes previously identified as cross-correlation and HT production are at maximum. We note that the HT basis vector is strongly aligned with the pure $H^+$ direction in the mass spectrum, although of course it contains other mass peaks as well. The orthogonal vector which emphasizes the cross-correlation of the UV and control pulse is strongly aligned with the parent peak in the mass spectrum. The dominance of $H^+$ in the HT mass spectrum agrees with reports by other groups on time-resolved studies of the CHD isomerization.[47,58-60]

In Fig. 1(b) we show the time evolution of the parent peak (red line) and the HT production (blue line) with pump-control time delay. The formation of HT is suppressed most efficiently when the control pulse arrives approximately 50 fs after the excitation pulse. The maximum suppression is approximately 10%. The displacement between the cross-correlation peak and the dip in HT production suggests that increased ionization due to the cross correlation cannot by itself account for the dip in the HT production. We propose that the control pulse is moving population from the excited state into the ground state, thus preventing the wavepacket from reaching the region of the potential energy surface where it can lead to isomerization.

In a further experiment where we varied the polarization of the control pulse with respect to that of the photoexcitation pulse, we observed an effect of this variation on the HT suppression. For each polarization arrangement, we integrated the ion count during the suppression (0-100 fs). This integrated ion count is shown in Fig. 1(c) as a function of the angle between the direction of polarization vector of the UV pulse and the direction of the polarization vector of the control pulse. Figure 1(c) shows that the fractional change in HT suppression is reduced by a factor of two for the perpendicular arrangement, compared to that of the parallel arrangement.

In summary, there are two main observations: HT production is suppressed by a control pulse, with maximum suppression of approximately 10% when the control pulse is delayed 50 fs after UV excitation. Furthermore, this suppression depends on the relative polarization of the UV and control pulse, and is largest when these polarizations are parallel. This interpretation is supported by *ab initio* multiple spawning calculations modeling the experiment, as described in the next section.

**4. Theoretical Results**

To directly simulate the interaction between an external laser and molecule, we extended our *ab initio* multiple spawning (AIMS)[61] with an external field modification as described elsewhere in detail.[62] Briefly, the total wavefunction ansatz is written as

$$\psi(r,R,t) = \sum_{I,n_j} \sum_j C_j^{I,n_j}(t) \chi_j^{I,n_j}(R,t) n_j \phi^{I,n_j}(r;R) \tag{1}$$

where r and R refer to electronic and nuclear coordinates respectively, $\phi^{I,n_j}$ is the electronic wavefunction of the $I^{th}$ adiabatic state dressed by $n_j$ photons, $\chi_j^{I,n_j}(R,t)$ is the $j^{th}$ time-dependent trajectory basis function (TBF) propagating on the $I^{th}$ electronic state, and $C_j^{I,n_j}$ is the complex amplitude of the $j^{th}$ TBF. The energy of a dressed electronic state is given as the field-free energy for the electronic state plus the energy derived from the photon field, i.e. $V^{I,n_j} = V^I(R) + n_j \hbar \omega$. The TBFs are frozen Gaussian wavepackets,[61,64,65] whose centers evolve according to classical equations of motion on the dressed electronic state that they are associated with. The time evolution of the amplitude $C_j^{I,n_j}$ is determined by solving the time-dependent Schrödinger equation within the time evolving basis set. The Hamiltonian operator used in this case includes three parts: the light field, the molecule, and their interaction:

$$\begin{aligned} H &= H_{light} + H_{molecule} + H_{interaction} \\ &= (aa^+ + \frac{1}{2})\hbar\omega + H_{molecule} + \vec{\mu} \cdot \vec{\varepsilon}(t)(a+a^+) \end{aligned} \tag{2}$$

where a and $a^+$ are the annihilation and creation operators in the photon field, respectively, and $\omega$ is the laser frequency.

The dressed states can be viewed in a Floquet picture with energies shifted from the non-dressed states by integer multiples of the photon energy, as shown in Fig 2. The solid lines indicate the field-free PESs (only $S_0$ and $S_1$ are shown), and the dashed lines are the one-photon dressed states. New conical intersections form when dressed states with different photon indices intersect with the original states, as indicated by the blue dots in the figure. We describe the molecule-light interaction within the dipole approximation here, but of course it is possible to include higher order terms in Eq. 2 if desired. The spawning procedure in AIMS is crucial in that it allows new TBFs to be created when they are needed because of impending nonadiabatic effects (these may or may not be populated according to the solution of the complex amplitudes via the finite basis time dependent Schrödinger equation). For usual CIs, the spawning procedure is triggered by large nonadiabatic couplings between electronic states. Here, we also must describe the possibility of transfer between dressed states due to LICIs. We do this by monitoring the interaction matrix elements between dressed states (last term in Eq. 2). When these exceed a threshold, spawning to a new dressed state will be triggered. AIMS will still spawn new TBFs near the normal conical intersections (indicated by the red dot in the figure) as usual. Therefore, the population distribution difference before and after applying the external field can be used to simulate the change in the CHD/HT branching ratio observed in the experiments.

The excited state dynamics of CHD without the control pulse were simulated and details will be described in a future publication.[62] Using these dynamics as a reference, we performed simulations with the external laser field arriving at various delays after the photoexcitation, in steps of 10 fs, provided that there was still population left on the excited state. The external laser field is described by a Gaussian function in the time domain with FWHM of 70 fs and field strength of 0.04 a.u. (similar to the experiment). In our model, the external laser field is polarized along the transition dipole of the molecule at the FC geometry, as this is the most probable direction for the molecule to absorb the pump radiation. The dynamics is followed for 200 fs after photoexcitation. Population transferred to the ground state by the external laser field was classified as CHD or HT according to its geometry at the end of the simulation. Fig. 3 shows the population transferred to the ground state by the laser and the branching between CHD and HT at the end of the simulations. The LICIs created by the 800 nm laser are most efficient from 30 to 70 fs, accounting for ~10% of the total population. After ~110 fs, there is no population transferred by the external laser field in our simulation. This is expected because most of the population has quenched to the ground electronic state by this time – our simulation (in the absence of any control pulses) predicts an excited state lifetime of 108 fs. To compare the simulation to the experimental measurement, we took the difference between the population that produced CHD and that which produced HT, and convolved the result with the experimental resolution based on the duration of the UV excitation pulse and the control pulse, as shown in Fig. 4.

## 5. Discussion

The simulations suggest that the coupling between $S_0$ and $S_1$ is most efficient at 50 fs where the population is transferred to the CHD ground state. This might be initially somewhat surprising since the established picture of CHD photochemistry has the molecule transferring from the bright 1B state to a dark 2A state before reaching the conical intersection which leads to the ground state. One might therefore think that the control laser field would not be able to promote population transfer once the molecule reached the dark 2A state. However, the key point is that the CHD molecule quickly loses its point group symmetry after excitation through out-of-plane motion of hydrogen atoms. Thus, the symmetry labels are no longer valid, the 1B and 2A states are mixed, and there is sufficient $S_1/S_0$ oscillator strength for the control pulse to promote optical transitions to the ground state. In general, the calculation accurately reproduces the experimental measurement magnitude and timing. This indicates that population transfer at the LICI is the major source of the control realized in the experiment.

If we assume that the control is a result of dipole coupling, as is indicated to be sufficient for a description of the experiment from the calculations, then the polarization findings provide clues to the evolving geometry of the molecule. As expected from the ultrafast excited state dynamics (and the relatively slow molecular reorientation), the observed polarization dependence suggests that the transition dipole moment between $S_1$ and $S_0$ at the control point is oriented approximately in the same direction as the transition dipole moment at the excitation point that is selected by the UV polarization. However, even for perpendicular polarization of the control pulse relative to the pump, HT production can still be suppressed; half of the population remains compared to the parallel polarization. This suggests that the transition dipole moment has reoriented somewhat compared to its original direction at the FC geometry. This can be rationalized by the rapid out-of-plane distortions experienced by CHD after photoexcitation.

## 6. Conclusion

In conclusion, we find that a strong infrared laser field (800nm) can control the photo-induced 1,3-cyclohexadiene to 1,3,5-hexatriene isomerization reaction. The field suppresses isomerization most efficiently if it is applied 50 fs after the initial photoabsorption. This suppression depends on the relative polarization of the UV and control pulse, and is greatest when the polarizations are parallel. At this delay the molecule is still in the electronically excited state, in the vicinity of a state character changing conical intersection that mediates this structural transition. The excited state is also in one-photon resonance with the ground state at this time. A modified *ab initio* multiple spawning method was developed which shows that the control laser induces a resonant coupling between the excited 2A/1B state and the ground 1A state. Our simulation shows that this is equivalent to a new light-induced conical intersection centered at the points of resonance between the ground and excited Born-Oppenheimer surfaces. The control pulse therefore diverts the molecule back to the ground state through a LICI before it can reach the conical intersection which would otherwise promote population transfer to the ground state. Since the geometry around the LICI in this case is more like the CHD reactant, this decreases the yield of HT photoproduct. The agreement between theory and experiment with respect to control pulse timing and product suppression validates this interpretation.

This control method probes the transient structure of a molecule in the vicinity of a conical intersection, and so it can be a useful tool for tracking vibrational wave packets during intramolecular processes. As a molecule evolves, the optimal control pulse should shift in wavelength, pulse delay, and polarization. We plan to explore these effects in the future.

ACKNOWLEDGMENT This experiment was supported by the National Science Foundation. Theory and simulation work was supported through the PULSE Institute by the U.S. Department of Energy Office of Basic Energy Sciences.

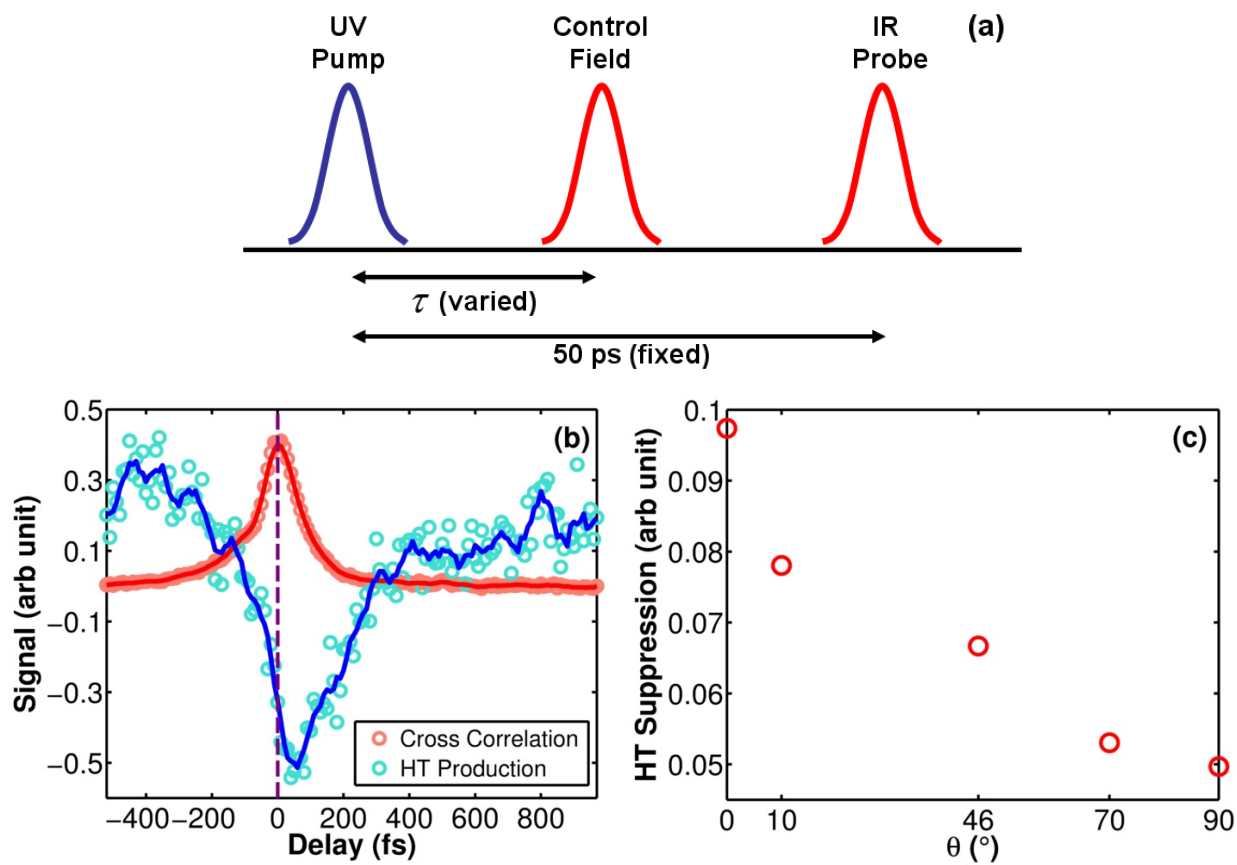

**Figure 1 (a)** Excitation scheme. The relative timing between the UV pump pulse and the IR fragmentation pulse was fixed at 50 ps. The delay of the control pulse with respect to the UV pulse, labeled τ, was scanned in the range between -500 fs and +900 fs in 10 fs steps. **(b)** When the UV excitation pulse and the control pulse overlap in time, all peaks in the ion-TOF mass spectrum increase (cross-correlation, red). Fragments associated with HT production (blue) show a visible suppression in their ion count at a delay of 50 fs. The displacement between these two features indicates that the control field drives population from the excited state into the ground state, thus suppressing the HT production. The solid lines are smoothed data by averaging five consecutive points of the raw data (dots) **(c)** Polarization dependence of the peak HT suppression displayed as a fraction of the total HT signal, and plotted vs. the angle between the polarization of the UV pulse and control pulse (θ). The ion count during the suppression (0-100 fs) was integrated and then divided by the baseline ion count. The fractional change in HT suppression is about 10% at θ = 0° and falls to 5% at θ = 90°.

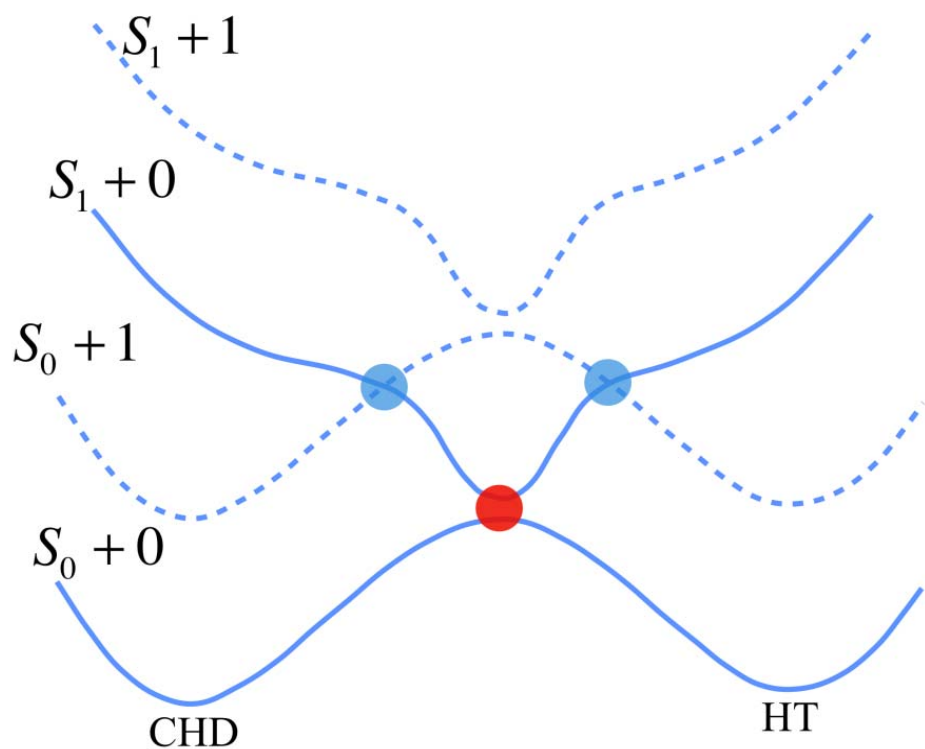

**Figure 2** Schematic representation showing the dressed states and a conical intersection in the presence of an external field. The solid lines are the bare energy levels (only $S_0$ and $S_1$ are shown). The dashed lines are states dressed by one photon. The laser-induced conical intersections are indicated by the blue dots, and the molecular field-free conical intersection is indicated by the red dot.

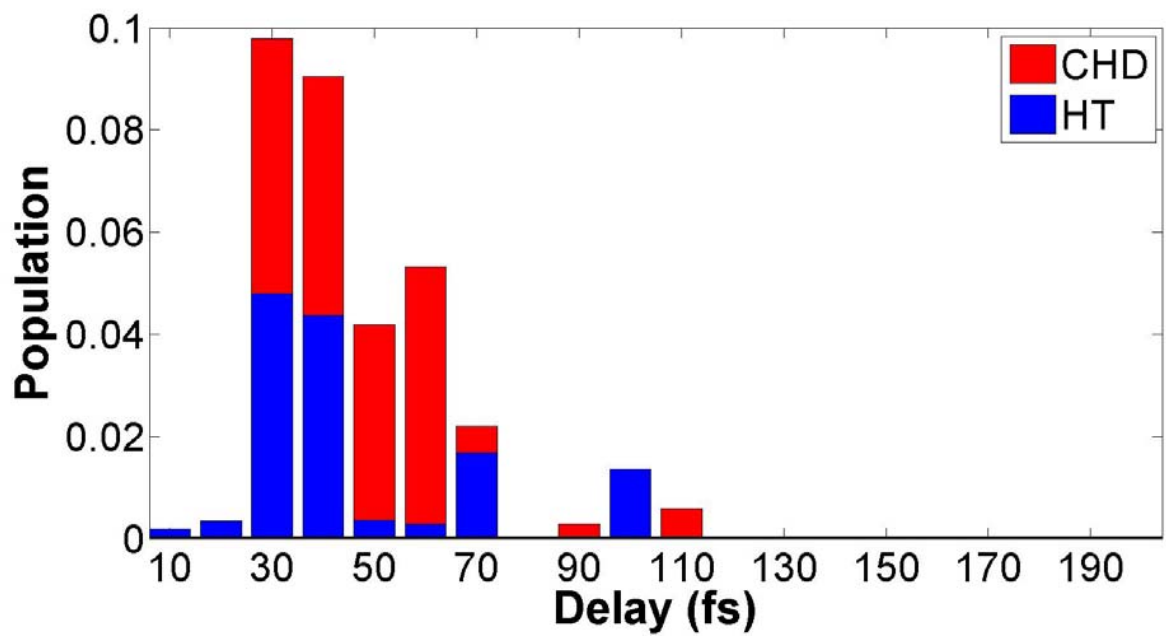

**Figure 3** Population of CHD and HT transferred to the ground state by the control field. The x axis is the time delay between the pump pulse and control pulse. The y axis is the population transfer to the ground state by the control pulse. The red (blue) bar indicates the transferred population that relaxes to the CHD (HT) geometry after evolution on the ground state.

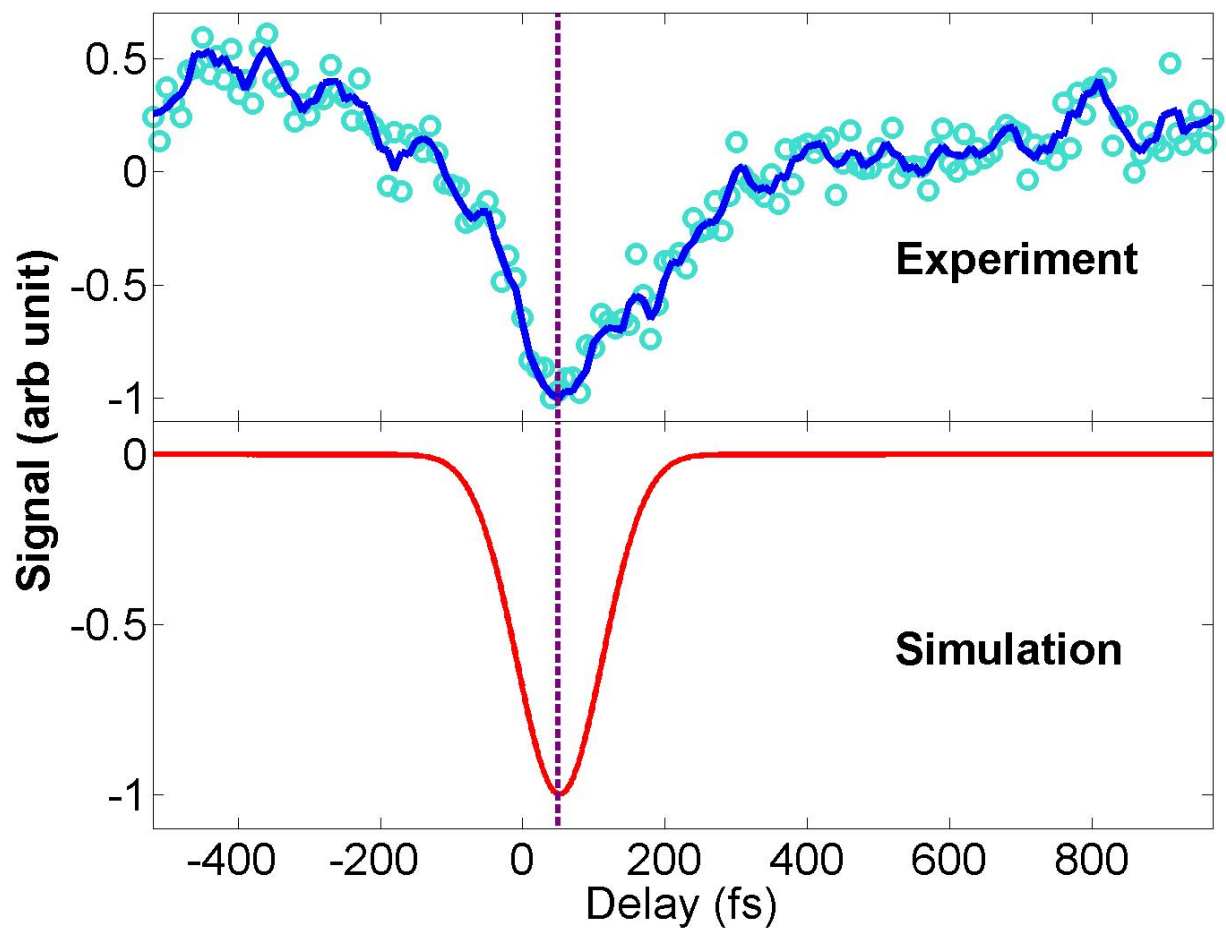

**Figure 4** Comparison between experimental (upper) and simulated (lower) signals for the branching ratio change induced by the external control field. The vertical line indicates the time delay when the control is maximal in the simulation, approximately 50 fs after the initial excitation.